\pdfoutput=1
\documentclass{article}
\usepackage{spconf,amsmath,graphicx}
\usepackage{algorithmic}
\usepackage{graphicx}
\usepackage{textcomp}
\usepackage{xcolor}
\usepackage{siunitx}
\usepackage{amsfonts}
\usepackage[symbol]{footmisc}

\usepackage{tabularx}
\usepackage{pifont}
\usepackage{booktabs}

\usepackage{hyperref}
\usepackage[capitalize]{cleveref}
\usepackage{tikz}
\usepackage{pgfplots}
\pgfplotsset{compat=1.18}
\usepackage{bm}
\usepackage{balance}
\usepackage{upgreek}
\usepackage{multirow} 
\usetikzlibrary{positioning, fit, calc}
\usepackage[acronym, shortcuts, toc,nonumberlist]{glossaries}

\usepackage{bbm}
\def\x{{\mathbf x}}
\def\L{{\cal L}}

\def\zspec{{z_{kt}^{\text{vM}}}}
\def\zspat{{z_{ltf}^{\text{cAC}}}}

\title{Loose coupling of spectral and spatial models for multi-channel diarization and enhancement of meetings in dynamic environments}
\newcommand{\upb}{$^1$}
\newcommand{\ntt}{$^2$}
\name{\tab{Adrian Meise\sthanks{Authors contributed equally}\upb, Tobias Cord-Landwehr$^*$\upb, Christoph Boeddeker\upb, Marc Delcroix\ntt,\\ Tomohiro Nakatani\ntt, Reinhold Haeb-Umbach\upb}}
\address{\upb Paderborn University, Germany \quad \ntt NTT, Inc., Japan}
\usepackage{xcolor}
\definecolor{magenta}{RGB}{138, 27, 97}
\definecolor{lightblue}{RGB}{0, 159, 223}
\definecolor{green}{RGB}{0, 155,119}
\definecolor{lightgreen}{RGB}{132, 189,0}
\definecolor{greenyellow}{RGB}{208, 223,0}

\definecolor{UPB_Ultrablau}{RGB}{0, 37, 170} 
\definecolor{UPB_Arktisblau}{RGB}{80, 209, 209}
\definecolor{UPB_Meerblau}{RGB}{35, 169, 201}

\definecolor{UPB_Himmelblau}{RGB}{10, 117, 196}
\definecolor{UPB_Saphirblau}{RGB}{24, 28, 98}
\definecolor{UPB_Irisviolett}{RGB}{126, 63, 168}
\definecolor{UPB_Fuchsiarot}{RGB}{193, 56, 160}
\definecolor{UPB_Granatpink}{RGB}{239, 58, 132}
\definecolor{UPB_Limettengruen}{RGB}{172, 234, 22}

\definecolor{green_teacher}{HTML}{8EA604}
\definecolor{yellow_student}{HTML}{EC9F05}
\usepackage{tikz}
\usetikzlibrary{arrows}
\usetikzlibrary{backgrounds}
\usetikzlibrary{fit}
\usetikzlibrary{positioning}
\usetikzlibrary{calc}
\usetikzlibrary{decorations.pathreplacing}
\usetikzlibrary{shapes.multipart}

\usetikzlibrary{shapes.misc}

\usepackage{tikzpagenodes}
\usetikzlibrary{tikzmark}
\usetikzlibrary{arrows.meta}  

\usepackage{pgfplots}
\pgfplotsset{compat=1.13}

\tikzset{
    >=stealth,
    block/.style={rectangle,draw=black!100,fill=UPB_Arktisblau!50,minimum size=3em,text height=1.5ex,line width=0.1em,text depth=.25ex},
    branch/.style={circle,draw=black,fill=black,minimum size=0.25em,inner sep=0pt},
    arrow/.style={->, line width=0.1em},
    line/.style={-, line width=0.1em},
    reverse arrow/.style={<-,shorten <=0.1em},
}

\begin{document}
\ninept
\setlength{\abovedisplayskip}{5pt}
\setlength{\belowdisplayskip}{5pt}
\setlength{\textfloatsep}{5pt plus 0.0pt minus 0.0pt}
\setlength{\floatsep}{5pt plus 0.0pt minus 2.0pt}

\def\x{{\mathbf x}}
\def\L{{\cal L}}
\newcommand{\cmark}{\text{\ding{51}}}
\newcommand{\xmark}{\text{\ding{55}}}

\newcommand{\fromtcl}[1]{ {\color{purple}  #1}}
\newcommand{\fromcb}[1]{ {\color{orange}  #1}}
\newcommand{\fromrh}[1]{ {\color{red}  #1}}

\newcommand{\mrow}[2][2]{\multirow{#1}[2]{*}{\begin{tabular}{@{}c@{}}#2\end{tabular}}}  
\newcommand{\tab}[2][c]{\begin{tabular}{@{}#1@{}}#2\end{tabular}}   

\newcolumntype{H}{>{\setbox0=\hbox\bgroup}c<{\egroup}@{}}   

\glsdisablehyper    
\newacronym{cACGMM}{cACGMM}{complex Angular Central Gaussian Mixture Model}
\newacronym{GSS}{GSS}{Guided Source Separation}
\newacronym{ASR}{ASR}{Automatic Speech Recognition}
\newacronym{vMFMM}{vMFMM}{{von-Mises-Fisher Mixture Model}}
\newacronym{vMFcACGMM}{vMFcACGMM}{von-Mises-Fisher complex Angular Central Gaussian Mixture Model}
\newacronym{VAD}{VAD}{Voice Activity Detection}
\newacronym{DER}{DER}{Diarization Error Rate}
\newacronym{EER}{EER}{Equal Error Rate}
\newacronym{EM}{EM}{Expectation-Maximization}
\newacronym{WER}{WER}{Word Error Rate}
\newacronym{cpWER}{cpWER}{concatenated minimum-permutation Word Error Rate}
\newacronym{SLR}{SLR}{Segment Level Reassignment}
\newacronym{EEND}{EEND}{End-to-End Neural Diarization}
\newacronym{GSS}{GSS}{Guided Source Separation}
\newacronym{SSND}{SSND}{Speaker Separation via Neural Diarization}
\maketitle
\begin{abstract}
Sound capture by microphone arrays opens the possibility to exploit spatial, in addition to spectral, information for diarization and signal enhancement, two important tasks in meeting transcription. However, there is no one-to-one mapping of positions in space to speakers if speakers move. Here, we address this by proposing a novel joint spatial and spectral mixture model, whose two submodels are loosely coupled by modeling the relationship between speaker and position index probabilistically. Thus, spatial and spectral information can be jointly exploited, while at the same time allowing for speakers speaking from different positions. Experiments on the LibriCSS data set with simulated speaker position changes show great improvements over tightly coupled subsystems.
\end{abstract}
\section{Introduction}
\label{sec:intro}
Automatic meeting transcription is commonly understood to comprise the tasks of diarization, answering the question of who speaks when, of speech enhancement, which includes the separation of overlapped speech, and of \gls{ASR}. 

 Diarization, which consists of the subtasks segmentation and speaker (re-)identification, started out heavily relying on spatial information in the context of meeting scenarios \cite{pardo2006speaker, araki2008doa, hager2008handling}. 
 Following that, more recent models mostly employed purely spectral approaches:
 Speaker embedding vectors, which capture a spectral profile of a speaker, are extracted from segments of speech and are then clustered to obtain the diarization result \cite{19_snyder_xvectors}, or are directly used to perform frame-level activity classification as in  \gls{EEND} techniques \cite{fujita19_eend, horiguchi20_eda_eend}.
 A combination of both approaches, as proposed in \cite{21_kinoshita_eend_vc}, can overcome the limitations of either system and is the basis of the widely used tools Pyannote and Diarizen \cite{23_plaquet_pyannote, 25_han_diarizen}. 
 While current systems again work on incorporating  multi-channel information\cite{22_horiguchi_mc_eend, 22_Zheng_tdoa_aug_dia, 24_ustc_chime8_dia}, these systems still primarily focus on the spectral information of a single channel and use the spatial cues as auxiliary information.

If the signals are recorded by a microphone array, spatial information can also be exploited to improve the speech enhancement quality. 
Array processing is well-established in the speech enhancement and source separation field, where both model-based, neural, and hybrid approaches have demonstrated significant performance gains over single-channel solutions \cite{spmag2024}. 
%
In \cite{boeddeker2022initialization}, a \gls{cACGMM} is employed, which models the spatial properties of the signal. Here, a mixture component corresponds to a position in space that is assumed to uniquely identify a speaker.
The aforementioned cACGMM is also the model underlying \gls{GSS} \cite{18_boeddeker_gss}. \Gls{GSS} takes the speech activity information of a preceding diarization stage as prior information for \gls{EM}-based model parameter estimation: The prior probabilities of the mixture components are clamped to zero if the preceding diarization indicates that the associated speaker is silent. The estimated affiliation values are the sought-after speaker activity probabilities at time-frequency (tf) resolution. They are then used for source extraction, either by masking or by beamforming. 

Instead of cascading diarization and enhancement, 
spatial information can also be leveraged in joint diarization and enhancement/separation approaches, where diarization and separation are coupled to benefit each other. 
This introduces additional challenges since voice characteristics need to be associated with positions that may change over the course of the recording.

An example of such an integration is \gls{SSND} \cite{24_taherian_ssnd}, which 
uses a multi-channel \gls{EEND} system to estimate short-time speaker embeddings, which are then processed jointly with the audio signal in a subsequent separation stage. 

A purely model-based approach to joint diarization and enhancement has been proposed in \cite{cord2025tightintegration}, where a spectral mixture model utilizing short-time speaker embeddings has been integrated with a spatial mixture model: they share a common latent source activity variable. Estimating its posterior probability gives tf-masks that incorporate both spectral and spatial information.

However, this method for combining spectral and spatial information performs a tight integration of the spatial and spectral models, i.e., it is based on the assumption that there is a one-to-one mapping of speaker location to speaker identity. 
Put differently, an underlying, crucial assumption is that speakers do not move. The goal of this paper is to relax this assumption and allow that the same speaker may speak from different positions in space. To achieve this, we propose a novel spectral-spatial mixture model, which we call a loosely coupled model.
In this model, the tight coupling of the original integrated model \cite{cord2025tightintegration} is released, and a probabilistic dependency between the spatial and the spectral mixture components is introduced, which in turn
allows mapping multiple spatial mixture components to a single spectral mixture component. Being model-based, the system does not require a training stage. This makes it operate on arbitrary microphone array configurations.

In the next section we first briefly describe the tightly integrated model-based spectral-spatial diarization system of \cite{cord2025tightintegration} and show in \cref{sec:loose_coupling} how it can be relaxed to a loose coupling. The key component is the conditional probability of a spatial mixture component, given the spectral one. An important aspect is how to estimate tf-masks for source extraction in this model, to which we present a solution in \cref{sec:mask_estimation}. Experiments on the LibriCSS data set in \cref{sec:evaluation} with simulated speaker position changes demonstrate the greatly increased insensitivity to speaker position changes compared to a tight integration of spectral and spatial diarization components. 

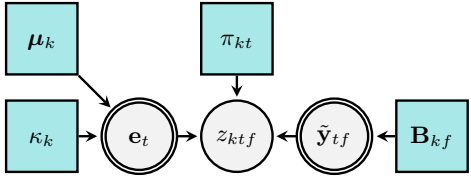
\begin{figure}
    \centering
    \begin{tikzpicture}
[
		node distance=1em,
		remember picture,
	]
            \node (r) [block] {$\boldsymbol \mu_{k}$};
		\node (kappa) [block, below=of r] {$\kappa_{k}$};
		\node (mix) [circle, minimum size=3em, text height=1.5ex, draw=black, line width=0.1em, fill=gray!10, double, right=of kappa] {$\mathbf{e}_t$};
            \node (aff) [circle, draw=black, minimum size=3em, text height=1.5ex, line width=0.1em,fill=gray!10, right=of mix] {$z_{ktf}$};
		\node (pi) [block, above=of aff] {$\pi_{kt}$};
            \node (obs) [circle, draw=black, double, minimum size=3em, text height=1.5ex, line width=0.1em, fill=gray!10, double, right=of aff] {$\tilde{\mathbf{y}}_{tf }$};		
            \node (sigma) [block, right=of obs] {$\mathbf{B}_{kf}$};

            \draw[arrow] (r) -- ($(mix.north west) + (-1pt, 1pt)$);
            \draw[arrow] (kappa) -- ($(mix.west) + (-2pt, 0pt)$);
            \draw[arrow] (mix) -- ($(aff.west) + (-1pt, 0pt)$);
            \draw[arrow] (obs) -- ($(aff.east) + (1pt, 0pt)$);
            \draw[arrow] (pi) -- ($(aff.north) + (0pt, 1pt)$);
            \draw[arrow] (sigma) -- ($(obs.east) + (2pt, 0pt)$);
\end{tikzpicture}
    \caption{Graphical model of the tight integration from \cite{cord2025tightintegration}. The spectral model (left) and the spatial model (right) are coupled through the common latent variable $z_{ktf}$.}
    \label{fig:tight_integration}
\end{figure}

\section{Mixture Model-based integration of separation \& diarization}
\label{sec:tight_integration}
In \cite{cord2025tightintegration}, diarization and speech separation were jointly addressed by modeling them by two statistical mixture models that are coupled through a common latent variable $\mathcal {Z}$ representing the speech activity of each speaker per tf-bin. This mixture model aims to estimate these categorically distributed latent variables $z_{ktf}$ given a set of observations $\mathcal{O}=\{\mathcal{O}_t~~\forall~t\}$.
Given a meeting recording with $C$ microphones, these observations are $\mathcal{O}_t=\{\{\tilde{\mathbf{y}}_{tf}\}_{f=1}^F,\mathbf{e}_t\}$. Here, $\tilde{\mathbf{y}}_{tf} \in \mathbb{C}^C$  denote the length-normalized multi-channel STFT features of the meeting with time frame index $t$, frequency bin index $f$, and the frame-level speaker embeddings $\mathbf{e}_t$ are obtained with a pre-trained speaker embedding extractor, such as \cite{cord2024geodesic}. The first captures spatial, and the second spectral information in the joint model
\begin{align}
\label{eq:obs_tight}
   p(\mathcal{O}_t)=\prod_f \sum_k  \pi_{kt} p^{\text{VMF}}(\mathbf{e}_t|z_{ktf}=1) p^{\text{cACG}}(\tilde{\mathbf{y}}_{tf}|z_{ktf}=1) ,
\end{align}
respectively. $\pi_{kt}$ is the time-dependent prior probability of each speaker in the meeting as introduced in \cite{16_ito_cacg_separation}.
For the integrated model, a \gls{vMFMM} \cite{05_banerjee_vmf} and a \gls{cACGMM} \cite{16_ito_cacg_separation} with respective component distributions $p^{\text{VMF}}(\cdot)$ \cite{53_fisher_vmf} and $p^{\text{cACG}} (\cdot)$ \cite{87_tyler_cacg} are combined to perform diarization and separation jointly, by estimating the latent variables $z_{ktf}$ of each speaker $k$ with the \gls{EM} algorithm. For the remainder of this work, the latent variables are denoted without the condition of \cref{eq:obs_tight} due to space constraints. The vMF distribution is specified by an average orientation $\boldsymbol \mu$ and the concentration parameter $\kappa$, which can be interpreted as the average speaker embedding and embedding uncertainty of a speaker in a meeting, while the cACG distribution is determined by the frequency-dependent spatial covariance matrix $\mathbf B_f$, which models the spatial cues of a location with active speech. 

The posterior probabilities $p(z_{ktf} | \mathcal{O}_t)$ then serve as mask estimates for each tf-bin and speaker $k$ and are used to reconstruct the individual utterances of the recording through beamforming. The graphical model of this tightly integrated model, i.e., a model using joint latent variables,  is visualized in \cref{fig:tight_integration}.

While the spatial mixture model is able to assign each tf-bin to a mixture component in this setup, the spectral mixture model attributes all frequency bins $f \in \{1,\dots F\}$ of a given time frame $t$ to the same class. This automatically introduces a mismatch in time frames, where a speaker is not active in all frequencies.
Further note that because a mixture component represents a speaker in the spectral model and a location in the spatial model, the tight coupling of the two models via the shared latent variable induces the assumption that there is a one-to-one mapping between speaker and location.
This prevents the modeling of moving speakers, because if a speaker speaks from different positions, multiple spatial components need to be mapped to the same spectral mixture component. 
Therefore, the tight coupling is by design incapable of tracking position changes.

\section{Loose coupling of mixture models}
\label{sec:loose_coupling}

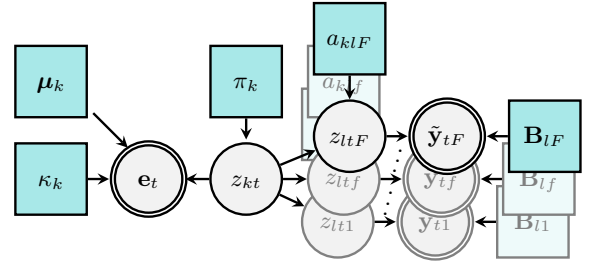
\begin{figure}
    \centering
    \begin{tikzpicture}[
		node distance=1em,
		remember picture,
	]
     \pgfdeclarelayer{fg}
      \pgfdeclarelayer{bg}
  \pgfsetlayers{bg,main,fg}
		\node (r) [block] {$\boldsymbol \mu_{k}$};
		\node (kappa) [block, below=of r] {$\kappa_{k}$};
		\node (mix) [circle, minimum size=3em, text height=1.5ex, draw=black, line width=0.1em, fill=gray!10, double, right=of kappa] {$\mathbf{e}_t$};
            \node (aff_spec) [circle, draw=black, minimum size=3em, text height=1.5ex, line width=0.1em,fill=gray!10, right=of mix] {$z_{kt}$};
		\node (pi) [block, above=of aff_spec] {$\pi_{k}$};
		\node (aff_spat)  [circle, draw=gray, text=gray, minimum size=3em, text height=1.5ex, line width=0.1em,fill=gray!10, right=of aff_spec] {$z_{ltf}$};
            \node (coupling) [block, draw=gray, fill=UPB_Arktisblau!10,text=gray, above=of aff_spat] {$a_{klf}$};
            \node (obs) [circle, draw=gray, text=gray, minimum size=3em, text height=1.5ex, line width=0.1em, fill=gray!10, double, right=of aff_spat] {$\tilde{\mathbf{y}}_{tf }$};		
            \node (sigma) [block,  fill=UPB_Arktisblau!10, draw=gray, text=gray, right=of obs] {$\mathbf{B}_{lf}$};   
            \begin{pgfonlayer}{bg}
                \node (coupling1) [block, draw=gray, fill=UPB_Arktisblau!10,text=gray] at ($(coupling.south) + (-0.25em, -0.25em)$) {$a_{kl1}$};
                \node (aff_spat1)  [circle, draw=gray, minimum size=3em,text=gray, text height=1.5ex, line width=0.1em,fill=gray!10] at ($(aff_spat.south) + (-0.25em, -0.25em)$) {$z_{lt1}$};
                \node (obs1) [circle, minimum size=3em, text height=1.5ex,text=gray, draw=gray, line width=0.1em, fill=gray!10, double]  at ($(obs.south) + (-0.25em, -0.25em)$)  {$\tilde{\mathbf{y}}_{t1}$};		
                \node (sigma1) [block, fill=UPB_Arktisblau!10, draw=gray, text=gray]  at ($(sigma.south) + (-0.25em, -0.25em)$)  {$\mathbf{B}_{l1}$}; 
                \draw[arrow] (coupling1)  edge node{} (aff_spat1);
		  \draw[arrow] (sigma1) edge node {} ($(obs1.east) + (1pt, 0pt)$);
            \draw[arrow] (aff_spat1) edge node {} ($(obs1.west) + (-1pt, 0pt)$);
            \draw[arrow] (aff_spec)  edge node{} (aff_spat1);

            \end{pgfonlayer}
            \begin{pgfonlayer}{fg}
                \node (couplingF) [block] at ($(coupling.north) + (0.25em, 0.25em)$) {$a_{klF}$};
                \node (aff_spatF)  [circle, draw=black, minimum size=3em,text=black, text height=1.5ex, line width=0.1em,fill=gray!10] at ($(aff_spat.north) + (0.25em, 0.25em)$) {$z_{ltF}$};
                \node (obsF) [circle, minimum size=3em, text height=1.5ex,text=black, draw=black, line width=0.1em, fill=gray!10, double]  at ($(obs.north) + (0.25em, 0.25em)$)  {$\tilde{\mathbf{y}}_{tF}$};		
                \node (sigmaF) [block]  at ($(sigma.north) + (0.25em, 0.25em)$)  {$\mathbf{B}_{lF}$};
                \draw[arrow] (couplingF)  edge node{} (aff_spatF);
                \draw[arrow] (sigmaF) edge node {} ($(obsF.east) + (1pt, 0pt)$);
            \draw[arrow] (aff_spatF) edge node {} ($(obsF.west) + (-1pt, 0pt)$);
            \draw[arrow] (aff_spec)  edge node{} (aff_spatF);
        
        \end{pgfonlayer}

            \node (dots1) [text=black, rotate=-15] at ($(aff_spat1.north east) + (1.1em, 0.em)$) {$\vdots$};
            \node (dots2) [text=black, rotate=-15] at ($(aff_spat.north east) + (1.25em, 0.em)$) {$\vdots$};

		\draw[arrow] ($(r.south east) + (2pt, 2pt)$) --  ($(mix.north west) + (2pt, 2pt)$);
		\draw[arrow] (kappa) edge node {} ($(mix.west) + (-1pt, 0pt)$);
          		
            \draw[arrow] (aff_spat) edge node {} ($(obs.west) + (-1pt, 0pt)$);
		\draw[arrow] (sigma) edge node {} ($(obs.east) + (1pt, 0pt)$);
		\draw[arrow] (pi) edge node {} ($(aff_spec.north) + (0pt, 1pt)$);

            \draw[arrow] (aff_spec) edge node {} (mix);
            \draw[arrow] (coupling)  edge node{} (aff_spat);
            \draw[arrow] (aff_spec)  edge node{} (aff_spat);


\end{tikzpicture}
    \caption{Graphical model of the loose coupling. The latent variable $z_{kt}$ of the spectral model serves as prior for the latent variables $z_{ltf}$ of the spatial models, which are fitted per frequency $f\in\{1,\dots, F\}$.}
    \label{fig:loose_coupling}
\end{figure}

To address both systematic issues of the tight integration, we propose to replace the above tight by a loose coupling of the \gls{vMFMM} and \gls{cACGMM}. This is done by reformulating the integration 
with one latent variable $\zspec$ for the spectral and another, $\zspat$, for the spatial mixture model, respectively. 

These two latent variables are assumed to be conditionally dependent with 
\begin{align}
    p(\zspat|\zspec) = a_{klf} \:\text{ with }\: \sum_l a_{klf} = 1, 
\end{align}
formulating the spectral, time-dependent affiliation $\zspec$ as prior for the spatial, time-frequency dependent affiliation $\zspat$. This allows the number of spectral VMF mixture components $K$ to be different from the number of spatial cACG mixture components $L$. The likelihood of the observed data is then given by 
\begin{align}
\begin{split}
    &p(\mathcal{O}_t) = 
     \sum_{k,l_{1}\dots l_{F}} \pi_kp(\mathbf{e}_t|\zspec)\prod_f a_{kl_{f}f} p(\tilde{\mathbf{y}}_{tf}|\zspat) \\
    &=\sum_k \pi_k p(\mathbf{e}_t|\zspec)\prod_f \sum_l a_{klf}p(\tilde{\mathbf{y}}_{tf} |\zspat) ,
\end{split}
\end{align}
where the coupling factors $a_{klf}$ can be viewed as the probability of a speaker being active from location $l$ given that speaker $k$ is active, and the prior $\pi_k$ of the spectral variables describes the probability of speaker $k$ being active.  The graphical model is visualized in \cref{fig:loose_coupling}.

In this way, the spectral model determines whether a speaker is active for a given time frame, while the spatial model can still portray inactivity for certain frequency bins at this time frame. 
At the same time, the loose coupling also allows for multiple spatial mixture components to correspond to the same speaker. Therefore, the model allows for a speaker being active from multiple locations over the course of the processed segments. Multiple spatial mixture components can also represent the same speaker by capturing strong reflections from walls, which makes the same speaker appear active from different positions. 

In order to fit this model to the observations $\mathcal{O}$,  the \gls{EM} algorithm is employed. As in \cite{cord2025tightintegration}, the M-step to estimate the distribution parameters $\boldsymbol \Theta= \{\boldsymbol{\mu}_k,\kappa_k, \pi_k, \mathbf{B}_{lf} ~~~\forall~klf\}$ remains unchanged compared to an individual \gls{vMFMM} or \gls{cACGMM}, with the difference being that the respective latent variables $\zspec$ and $\zspat$ are used.
For the E-step, the individual latent variables can be estimated by first obtaining the joint latent posterior 
\begin{align}
\begin{split}
        &\delta_{kltf}= p(\zspec,\zspat| \mathcal O_t, \boldsymbol \Theta) \\
        &= \frac{\pi_k p(\mathbf{e}_t|\zspec)\alpha_{klf}p(\tilde{\mathbf{y}}_{tf}|\zspat)}{p(\mathcal{O}_t)} \left(\prod_{f' \neq f}\sum_l\alpha_{klf'}p(\tilde{\mathbf{y}}_{tf'}|z_{ltf'}^{\text{cAC}})\right),
\end{split}
\end{align}
from which the estimates of the individual posteriors 
\begin{align}
    p(\zspec|\mathcal{O}_t) = \frac{1}{F}\sum_{l}\sum_f\delta_{kltf}, \quad p(\zspat | \mathcal O_t) = \sum_k \delta_{kltf}
\end{align}
can be obtained via marginalization.
The newly introduced coupling weights are estimated in the M-step by
\begin{align}
\label{eq:alpha}
    \alpha_{klf} = 
    \frac{\sum_t \delta_{kltf}}{\sum_t\sum_l\delta_{kltf}}.
\end{align}

\subsection{Mask estimation for loosely coupled models}
\label{sec:mask_estimation}
Contrary to the tight integration of \cref{sec:tight_integration}, the affiliations of the loosely coupled model cannot be directly interpreted as mask estimates for each speaker in the processed recording. The mixture model only contains location-specific affiliations $\zspat$ and speaker-specific diarization affiliations $\zspec$. 
For speech extraction, speaker-specific mask estimates $m_{ktf}$ are required in order to apply them for mask-based beamforming.
Since only the spatial model, which is dependent on $l$, depicts frequency-dependent information, marginalization over the locations of the joint posterior $\delta_{kltf}$ results in a constant solution over all frequencies. This is undesirable for the extraction process since it impairs the performance, especially in regions of multiple active speakers and strong background noise.

Instead, to make the estimation more reliable, we introduce heuristics for the weight calculation and the modification of the joint posterior. 
First, the probability that a speaker is active, given that the location is active for a single frequency $f$ 
\begin{align}
    \beta_{klf} = p(\zspec |\zspat, \mathcal{O}_t)
\end{align}
is estimated complementary to the coupling factors $a_{klf}$. 
Then, these values are used as a weighting factor for the joint posterior to obtain frequency-selective masks.
$\beta_{klf}$ is estimated by first removing the location 
corresponding to noise from the joint posterior 
 $\delta_{kltf}$.  This location is identified as the one carrying the highest overall activity, as was done in \cite{boeddeker2022initialization}, since speech sources carry only sparse activity.
Then, similar to \cref{eq:alpha}, the reduced posterior $\tilde{\delta}_{kltf}$ without noise component is used to estimate
\begin{align}
    \beta_{klf} = 
    \frac{\sum_t\tilde\delta_{kltf}}{\sum_t\sum_k\tilde{\delta}_{kltf}}
\end{align}
directly from the joint posterior.
To account for noise contributions in the remaining location components, $\beta_{klf}$ is then thresholded
at  
\begin{align}
  \tilde{\beta}_{klf} = \begin{cases}
    \beta_{klf} & \text{if } \beta_{klf} \geq \tau_{\text{th}} \\
      0 & \text{else}\\
   \end{cases}
\end{align}
with the threshold $\tau_{\text{th}}=0.55$
to only consider the relevant positions and not reflections or noise interference. 
Finally, the mask estimates 
\begin{align}
    m_{ktf} =  \sum_l\tilde{\beta}_{klf}\tilde{\delta}_{kltf}.
\end{align}
 are obtained by multiplying the mapping of location to speaker with the joint posterior before
  marginalization of the locations. 

\section{Evaluation}
\label{sec:evaluation}
We evaluate the model using the LibriCSS database \cite{20_Chen_libricss}, which consists of re-recorded, simulated meetings of \num{8} stationary speakers containing \SIrange{0}{40}{\percent} overlapping speech (0S -- OV40).
Here, the performance is evaluated both on the LibriCSS segments, a predefined split of the LibriCSS meetings at silence with an average segment duration of \SI{50}{\second}, and a concatenated version of the LibriCSS segments. 
For this version, 
two segments of the same session are concatenated such that each segment is used only once and the number of speakers common to both segments is maximized\footnote{\url{https://zenodo.org/records/17121135}}, resulting in an average segment duration of \SI{75}{\s}.

In addition, an evaluation set with speaker position changes is simulated for the concatenated segments by rotating all non-center microphone channels by two positions in the latter segment, effectively causing each speaker to move by \SI{120}{\degree}.
For this setup, which we call speaker relocation, on average, \num{3.2} active speakers 
are common to both segments.

For the initialization of the loose coupling, the joint posterior $\delta_{kltf}$ is initialized via an individual spectral and spatial initialization.
For the spectral initialization, a fusion-based scheme as proposed in \cite{cord2025tightintegration} is employed. First, an energy-based \gls{VAD} using minimum statistics is applied to run a 
k-Means clustering using $2K$ classes on the embeddings corresponding to speech. Then, a \gls{vMFMM} is fitted to the data, where the mixture components with the highest similarity are fused until $K$ clusters remain. These activity estimates for the speakers are then extended across the speech pauses detected by the VAD, obtaining $z_{kt}^0$. Alternatively, we also investigate using the diarization target as oracle initialization as a reference.
For the initialization of the spatial model, the clustering-based approach from \cite{boeddeker2022initialization} is used. 
Following that, a \gls{cACGMM} is fitted to the data for \num{30} iterations to stabilize the spatial initialization $z_{ltf}^0$.
We assume the number of speakers $N$ in a segment is known and choose $K=N$ spectral components and $L=2N+1$ spatial positions w.l.o.g., providing two positions for each speaker as well as an additional noise class.
The initial estimates of the spectral and spatial initialization are finally combined into the initial $\delta_{kltf}^0=z_{kt}^0 z_{ltf}^0$. Following that, the model is started with an M-step to obtain the parameters $\boldsymbol \Theta$. 

We compare the loose coupling with a \gls{cACGMM} from \cite{boeddeker2022initialization}
and the tight integration from \cite{cord2025tightintegration} as baselines, where each model is fitted for \num{100} EM iterations to ensure convergence. The detected activity is first smoothed, and duplicates of speakers are removed, which are identified by comparing the intersection of their activity patterns and their prototypical speaker representations $\boldsymbol{\mu}$. 
\begin{table*}[bht]
\centering
\vspace{-6pt}
\caption{cpWER of the tight integration and loosely coupled system on the concatenated LibriCSS segments for the \textit{static} (left/) and the \textit{speaker relocation} (/right) scenario. ``orc. init`` denotes an initialization with the oracle diarization.}
\vspace{3pt}
\label{tab:concat_performance}
\sisetup{round-mode=places,round-precision=1}
\begin{tabular}{l c S@{~/~}S S@{~/~}S S@{~/~}S S@{~/~}S S@{~/~}S S@{~/~}S | S@{~/~}S}
\toprule
    System & orc. init &\multicolumn{2}{c}{{0S}} & \multicolumn{2}{c}{{0L}} & \multicolumn{2}{c}{{OV10}} & \multicolumn{2}{c}{{OV20}} & \multicolumn{2}{c}{{OV30}} & \multicolumn{2}{c}{{OV40}} & \multicolumn{2}{c}{{avg.}}  \\
    \midrule
    cACGMM & \cmark & 4.22& 19.68 &4.10 & 22.97 &3.52 & 21.97 &4.56 & 24.18 & 7.86 & 28.26 &6.42 & 29.65 & 5.30 & 24.84 \\
    tight int. & \cmark & 4.23 & 17.16& 4.35 & 21.88 & 4.15 & 19.41 & 5.05 & 23.36 & 5.80 & 25.93& 5.93 & 27.79& 5.01 & 22.94 \\
    loose coupling & \cmark & 5.0  &  5.0  &  3.5  &  5.0  & 2.7  & 5.0  & 3.4   & 6.8   &  4.3   &  11.1  &   4.1  &  12.7  & 3.9 & 8.0 \\
    \midrule
    tight int. & - &6.90 & 22.25 & 4.66 & 34.29& 6.92 & 25.25 & 7.80 & 25.89& 9.11 & 28.17 & 9.38 & 30.10 & 7.71& 27.47\\
    loose coupling & - & 6.9 & 9.3  & 5.3 & 8.4  & 4.0  & 9.2  & 5.8 & 12.9  &  6.9 & 15.2  & 7.0  &  19.4 & 6.0  & 12.9 \\
     \bottomrule
\end{tabular}
\vspace{-10pt}
\end{table*}

\begin{table}
\vspace{-6pt} 
\caption{cpWER of the tight integration and loosely coupled system on individual LibriCSS segments. 
``orc.`` denotes an initialization with the oracle diarization.\strut}
\vspace{3pt}
\label{tab:segment_level_performance}
\setlength{\tabcolsep}{1pt}
\centering
\begin{tabular}{l c@{~~~}S@{~~~~}S@{~~~}S@{~~~}S@{~~~}S@{~~~}S@{~~~}S@{~~~}S}
\toprule
System & orc. &  {0S} & {0L} & {OV10} & {OV20} & {OV30} & {OV40} & {avg.}\\
\midrule
    tight int. & \cmark &  4.8 & 3.8 & 3.1 & 4.2 & 5.0 & 4.9 & 4.3\\
    loose coup. & \cmark & 4.7 & 2.9  & 3.4   & 3.7  & 4.3  & 4.6 & 4.0 \\ 
    \midrule
    tight int. & - & 4.3 &5.9 & 3.9 & 4.9 & 6.5 & 6.8 & 5.4 \\ 
    loose coup. & - & 5.8 & 4.5 & 5.1 & 4.9 & 6.7 & 7.1 & 5.8 \\
     \bottomrule
\end{tabular}
\end{table}

Then, a masked-based MVDR beamformer \cite{09_souden_mvdr} is applied, and the utterances are transcribed with a Conformer-based \gls{ASR} model from the Nemo toolkit \cite{23_rekesh_nemo_asr, nemo_conformer_large}. The performance is compared in terms of the \gls{cpWER} \cite{chime6} calculated with the MeetEval toolkit \cite{MeetEval23}.

\subsection{Evaluation in the presence of speaker position changes}
\label{sec:relocation}
First, the performance on the concatenated LibriCSS segments without microphone channel rotation is compared against the \textit{relocation} setup with channel rotation
 to directly assess the influence of position changes. The results are shown in \cref{tab:concat_performance}.

For the static scenario, the loose coupling consistently shows an improvement over the 
tight integration, both for an oracle diarization and a non-oracle initialization of the spectral model.
For an oracle initialization, the loose coupling shows an improvement of \SI{1.1}{\percent} absolute in terms of \gls{cpWER}, while this improvement increases to \SI{1.7}{\percent} for a non-oracle initialization.
However, both the tight integration and the loose coupling increasingly lose performance for higher overlap subsets with a non-oracle initialization.
This effect is likely to come from the higher complexity of capturing all speakers active in a segment, since speakers with little or no single-speaker activity become more likely to appear for higher overlaps. 

When comparing the static results (left) with the \textit{relocation} subset (right) in \cref{tab:concat_performance}, the main advantage of the loose coupling becomes apparent.
Here, the tight integration fails to stabilize the system in the case of moving speakers. Even for an oracle initialization, the model is incapable of maintaining the diarization estimate, resulting in an increase in \gls{cpWER} from \SIrange{5.0}{22.9}{\percent}.  
This is almost as bad as when solely using a \gls{cACGMM}, which by design is a fully spatial model and incapable of tracking any position changes. 
Compared to that, the loose coupling outperforms the tight integration by \SI{14.9}{\percent} and \SI{14.6}{\percent} absolute for an oracle and non-oracle initialization, respectively.
With corresponding average \glspl{cpWER} of \SI{8.0}{\percent} and \SI{12.9}{\percent}, this still indicates a large performance loss compared to the static scenario, even if not as severe as for the tight integration.
However, when referring to the individual overlap subsets, it turns out that the performance loss of the loosely coupled model predominantly occurs for the high-overlap subsets.

This is similar to the results of the frame-level speaker embeddings of \cite{cord2024geodesic}, where a \gls{DER} of more than \SI{30}{\percent} occurs when scoring only overlapping speech. 
This implies that the embedding quality is significantly lower for regions of overlapping speech.
Therefore, we conjecture that the worse embedding quality under overlapping speech, jointly with the increased complexity of moving speakers, leads to higher amounts of missed speakers and speaker swaps, which negatively impact the WER.
Still, the model proves able to stabilize the performance in case of changing speaker positions, especially for low overlap ratios. 

\subsection{Performance in a stationary environment}
To directly assess the loose coupling's performance compared to the tight integration from \cite{cord2025tightintegration}, the performance is also compared on the individual LibriCSS segments.  
\cref{tab:segment_level_performance} shows this comparison.
Similar to the concatenated LibriCSS segments, the loose coupling is able to show an improvement upon the tight integration for oracle initialization, which is likely to come from the frequency selectivity and higher
flexibility in terms of mapping multiple locations to the same speaker components.

Contrary to the concatenated setup, for a non-oracle initialization, the loose model incurs performance losses compared to the tight integration, resulting in a performance drop from \SIrange{5.4}{5.8}{\percent} \gls{cpWER}. 
Again, the performance loss is more prominent for overlapping speech. 
Since the loose coupling puts a high reliance on the spectral features by using them as a prior for the spatial model, 
in cases where the frame-wise embeddings or the spectral initialization become unreliable, the spatial model can not as easily compensate for this information as for the tight integration.
Further, the quality of the spectral initialization becomes more important and a limiting factor of the performance which matches the findings from \cref{sec:relocation}.
However, by using a better spectral initialization, as shown for the use of oracle information, the model can be guided to a better solution during the fitting process. 

As the loose coupling needs to estimate the coupling weights during the fitting process of the model, it benefits more from a higher amount of observations 
which can also help for a better spectral initialization. 
Therefore, 
the loose coupling still shows an improvement upon the tight integration for longer segments as discussed in \cref{sec:relocation}, even in a static scenario.

\section{Conclusion}
In this work, we proposed a statistical dependence for mapping spatial and spectral mixture models in the context of meeting processing and introduced a way to obtain mask estimates from this loosely coupled model.
We were able to demonstrate that the loosely coupled system outperforms a tight integration of both models in the case of an oracle diarization. In case of longer segments, as is the case for a concatenation of two LibriCSS segments, the loose coupling model also outperforms a tight integration for a non-oracle initialization.
In addition, the loosely coupled model proves able to handle speakers being active from multiple positions, whereas the tight integration is unable to resolve mismatching spatial and spectral cues.

Currently, the loosely coupled model still suffers from performance degradation as the degree of overlapping speech increases. In addition, it seemingly requires more observations to perform a stable fitting process compared to the tight integration. 
Therefore, future work will aim at incorporating spectral observations with higher stability, especially for overlapping speech, and the application of the loose coupling to recordings with real speaker movements to increase the model robustness and assess the performance for speaker trajectories.

\section{Acknowledgments}
Computational resources were provided by BMBF/NHR/PC2

\bibliographystyle{IEEEbib}
\balance
\bibliography{strings,refs}

@string{icassp = "Proc. ICASSP"}

@string{interspeech = "Proc. Interspeech"}

@string{chime = "Proc. CHiME"}

@string{iwaenc = "Proc. IEEE IWAENC"}

@string{asru = "Proc. ASRU"}

@string{ieee-tsp = "IEEE Trans. Signal Process."}

@string{eusipco = "Proc. EUSIPCO"}

@inproceedings{pardo2006speaker,
  title={Speaker diarization for multi-microphone meetings using only between-channel differences},
  author={Pardo, J. M. and Anguera, X. and Wooters, C.},
  booktitle={International Workshop on Machine Learning for Multimodal Interaction},
  pages={257--264},
  year={2006},
  organization={Springer}
}

@inproceedings{araki2008doa,
  title={A {DOA} based speaker diarization system for real meetings},
  author={Araki, S. and Fujimoto, M. and Ishizuka, K. and others},
  booktitle={Hands-Free Speech Communication and Microphone Arrays},
  pages={29--32},
  year={2008},
  organization={IEEE}
}

@inproceedings{hager2008handling,
  title={Handling speaker position changes in a meeting diarization system by combining {DOA} clustering and speaker identification},
  author={Hager, T. and Araki, S. and Ishizuka, K. and others},
  booktitle=iwaenc,
  volume={106},
  year={2008}
}

@inproceedings{cord2025tightintegration,
  title={Simultaneous Diarization and Separation of Meetings through the Integration of Statistical Mixture Models},
  author={Cord-Landwehr, Tobias and Boeddeker, Christoph and Haeb-Umbach, Reinhold},
  booktitle=icassp,
  pages={1--5},
  year={2025},
  organization={IEEE}
}

@INPROCEEDINGS{22_horiguchi_mc_eend,
  author={Horiguchi, Shota and Takashima, Yuki and García, Paola and Watanabe, Shinji and Kawaguchi, Yohei},
  booktitle=icassp, 
  title={Multi-Channel End-To-End Neural Diarization with Distributed Microphones}, 
  year={2022},
  volume={},
  number={},
  pages={7332-7336},
  doi={10.1109/ICASSP43922.2022.9746749}}

@article{24_ustc_chime8_dia,
  title={Incorporating spatial cues in modular speaker diarization for multi-channel multi-party meetings},
  author={Wang, Ruoyu and Niu, Shutong and Yang, Gaobin and Du, Jun and Qian, Shuangqing and Gao, Tian and Pan, Jia},
  journal={arXiv preprint arXiv:2409.16803},
  year={2024}
}

@INPROCEEDINGS{22_Zheng_tdoa_aug_dia,
  author={Zheng, Naijun and Li, Na and Yu, JianWei and Weng, Chao and Su, Dan and Liu, XunYing and Meng, Helen},
  booktitle=icassp, 
  title={Multi-Channel Speaker Diarization Using Spatial Features for Meetings}, 
  year={2022},
  volume={},
  number={},
  pages={7337-7341},
  doi={10.1109/ICASSP43922.2022.9747343}}

@inproceedings{cord2024geodesic,
  title={Geodesic interpolation of frame-wise speaker embeddings for the diarization of meeting scenarios},
  author={Cord-Landwehr, Tobias and Boeddeker, Christoph and Zoril{\u{a}}, C{\u{a}}t{\u{a}}lin and Doddipatla, Rama and Haeb-Umbach, Reinhold},
  pages={11886--11890},
  year={2024},
  booktitle=icassp,
  organization={IEEE}
}

@article{87_tyler_cacg,
  title={Statistical analysis for the angular central Gaussian distribution on the sphere},
  author={Tyler, David E},
  journal={Biometrika},
  volume={74},
  number={3},
  pages={579--589},
  year={1987},
  publisher={Oxford University Press}
}

@inproceedings{16_ito_cacg_separation,
  title={Complex angular central Gaussian mixture model for directional statistics in mask-based microphone array signal processing},
  author={Ito, Nobutaka and Araki, Shoko and Nakatani, Tomohiro},
  booktitle=eusipco,
  pages={1153--1157},
  year={2016}
}

@inproceedings{boeddeker2022initialization,
  title={An initialization scheme for meeting separation with spatial mixture models},
  author={Boeddeker, Christoph and Cord-Landwehr, Tobias and von Neumann, Thilo and Haeb-Umbach, Reinhold},
  booktitle=interspeech,
  year={2022}
}

@article{05_banerjee_vmf,
  author  = {Arindam Banerjee and Inderjit S. Dhillon and Joydeep Ghosh and Suvrit Sra},
  title   = {Clustering on the Unit Hypersphere using von Mises-Fisher  Distributions},
  journal = {Journal of Machine Learning Research},
  year    = {2005},
  volume  = {6},
  number  = {46},
  pages   = {1345--1382},
  url     = {http://jmlr.org/papers/v6/banerjee05a.html}
}

@inproceedings{20_Chen_libricss,
	author={Chen, Zhuo and Yoshioka, Takuya and Lu, Liang and Zhou, Tianyan and Meng, Zhong and Luo, Yi and Wu, Jian and Xiao, Xiong and Li, Jinyu},
	booktitle=icassp, 
	title={Continuous Speech Separation: Dataset and Analysis}, 
	year={2020},
	pages={7284-7288},
	doi={10.1109/ICASSP40776.2020.9053426}}

@inproceedings{18_boeddeker_gss,
  author={Christoph Boeddecker and Jens Heitkaemper and Joerg Schmalenstroeer and Lukas Drude and Jahn Heymann and Reinhold Haeb-Umbach},
  title={{Front-end processing for the CHiME-5 dinner party scenario}},
  year=2018,
  booktitle=chime,
  pages={35--40},
  doi={10.21437/CHiME.2018-8}
}

@inproceedings{23_rekesh_nemo_asr,
  title={Fast conformer with linearly scalable attention for efficient speech recognition},
  author={Rekesh, Dima and Koluguri, Nithin Rao and Kriman, Samuel and Majumdar, Somshubra and Noroozi, Vahid and Huang, He and Hrinchuk, Oleksii and Puvvada, Krishna and Kumar, Ankur and Balam, Jagadeesh and others},
  booktitle=asru,
  pages={1--8},
  year={2023},
}

@inproceedings{fujita19_eend,
  title     = {End-to-End Neural Speaker Diarization with Permutation-Free Objectives},
  author    = {Yusuke Fujita and Naoyuki Kanda and Shota Horiguchi and Kenji Nagamatsu and Shinji Watanabe},
  year      = {2019},
  booktitle = interspeech,
  pages     = {4300--4304},
  doi       = {10.21437/Interspeech.2019-2899},
  issn      = {2958-1796},
}

@inproceedings{25_han_diarizen,
  title={Leveraging self-supervised learning for speaker diarization},
  author={Han, Jiangyu and Landini, Federico and Rohdin, Johan and Silnova, Anna and Diez, Mireia and Burget, Luk{\'a}{\v{s}}},
  booktitle=icassp,
  year={2025}
}

@inproceedings{21_kinoshita_eend_vc,
  title={Integrating end-to-end neural and clustering-based diarization: Getting the best of both worlds},
  author={Kinoshita, K. and Delcroix, M. and Tawara, N.},
  booktitle=icassp,
  pages={7198--7202},
  year={2021},
  organization={IEEE}
}

@inproceedings{23_plaquet_pyannote,
  author={Alexis Plaquet and Hervé Bredin},
  title={{Powerset multi-class cross entropy loss for neural speaker diarization}},
  year=2023,
  booktitle=interspeech,
}

@inproceedings{horiguchi20_eda_eend,
  title     = {End-to-End Speaker Diarization for an Unknown Number of Speakers with Encoder-Decoder Based Attractors},
  author    = {Shota Horiguchi and Yusuke Fujita and Shinji Watanabe and Yawen Xue and Kenji Nagamatsu},
  year      = {2020},
  booktitle = interspeech,
  pages     = {269--273},
  doi       = {10.21437/Interspeech.2020-1022},
  issn      = {2958-1796},
}

@InProceedings{MeetEval23,
  author    = {von Neumann, Thilo and Boeddeker, Christoph and Delcroix, Marc and Haeb-Umbach, Reinhold},
  title     = {{MeetEval}: A Toolkit for Computation of Word Error Rates for Meeting Transcription Systems},
  year      = {2023},
  booktitle = chime,
  pages     = {27--32},
  doi       = {10.21437/CHiME.2023-6}
}

@ARTICLE{spmag2024,
  author={Haeb-Umbach, Reinhold and Nakatani, Tomohiro and Delcroix, Marc and Boeddeker, Christoph and Ochiai, Tsubasa},
  journal={IEEE Signal Processing Magazine}, 
  title={Microphone Array Signal Processing and Deep Learning for Speech Enhancement: Combining model-based and data-driven approaches to parameter estimation and filtering}, 
  year={2024},
  volume={41},
  number={6},
  pages={12-23},
  doi={10.1109/MSP.2024.3451653}
}

@inproceedings{chime6,
  title     = {CHiME-6 Challenge: Tackling Multispeaker Speech Recognition for Unsegmented Recordings},
  author    = {Shinji Watanabe and Michael Mandel and Jon Barker and Emmanuel Vincent and Ashish Arora and Xuankai Chang and Sanjeev Khudanpur and Vimal Manohar and Daniel Povey and Desh Raj and David Snyder and Aswin Shanmugam Subramanian and Jan Trmal and Bar Ben Yair and Christoph Boeddeker and Zhaoheng Ni and Yusuke Fujita and Shota Horiguchi and Naoyuki Kanda and Takuya Yoshioka and Neville Ryant},
  year      = {2020},
  booktitle = chime,
  pages     = {1--7},
  doi       = {10.21437/CHiME.2020-1},
}

@article{24_taherian_ssnd,
  title={Multi-channel conversational speaker separation via neural diarization},
  author={Taherian, Hassan and Wang, DeLiang},
  journal=ieee-tsp,
  volume={32},
  pages={2467--2476},
  year={2024},
  publisher={IEEE}
}

@INPROCEEDINGS{19_snyder_xvectors,
  author={Snyder, David and Garcia-Romero, Daniel and Sell, Gregory and McCree, Alan and Povey, Daniel and Khudanpur, Sanjeev},
  booktitle=icassp, 
  title={Speaker Recognition for Multi-speaker Conversations Using X-vectors}, 
  year={2019},
  volume={},
  number={},
  pages={5796-5800},
  doi={10.1109/ICASSP.2019.8683760}}

@article{09_souden_mvdr,
  title={On optimal frequency-domain multichannel linear filtering for noise reduction},
  author={Souden, Mehrez and Benesty, Jacob and Affes, Sofiene},
  journal=ieee-tsp,
  volume={18},
  number={2},
  pages={260--276},
  year={2009},
  publisher={IEEE}
}

@article{53_fisher_vmf,
author = {Fisher, Ronald Aylmer },
title = {Dispersion on a sphere},
journal = {Proceedings of the Royal Society of London. Series A. Mathematical and Physical Sciences},
volume = {217},
number = {1130},
pages = {295-305},
year = {1953},
doi = {10.1098/rspa.1953.0064},

URL = {https://royalsocietypublishing.org/doi/abs/10.1098/rspa.1953.0064},
eprint = {https://royalsocietypublishing.org/doi/pdf/10.1098/rspa.1953.0064}
,
}

@misc{nemo_conformer_large,
  author       = {Oleksii Kuchaiev},
  title        = {{Nemo pretrained model, nvidia/stt\_en\_conformer\_ctc\_large}},
  year         = {2022},
  publisher    = {Huggingface},
  url          = {{https://huggingface.co/nvidia/stt_en_conformer_ctc_large}}
}

\end{document}